\documentclass[aps,prd,preprintnumbers,showpacs,showkeys,nofootinbib,
superscriptaddress,fleqn,floatfix,tightenlines,10pt,
byrevtex]{revtex4-2}
\usepackage{amsmath,amsfonts,amssymb,amscd,amsxtra,amsthm}
\usepackage{appendix}
\usepackage[unicode]{hyperref}
\usepackage{graphicx,subfigure}
\graphicspath{{./Figures/}}
\usepackage{indentfirst}
\usepackage{float}
\usepackage{bm}
\usepackage{physics}
\usepackage{multirow}
\usepackage{makecell}
\usepackage{ltablex}% <-- added
\usepackage{siunitx}% <-- added

\usepackage[utf8]{inputenc}
\usepackage[normalem]{ulem} % \sout{old text} for strikeout
\usepackage[dvipsnames]{xcolor} % For blue in-text comments and
\graphicspath{{./Figures/}}
\usepackage{indentfirst}
\usepackage{float}
\usepackage[dvipsnames]{xcolor}

\newcommand{\Slash}[1]{\ooalign{\hfil/\hfil\crcr$#1$}}

\usepackage[compat=1.1.0]{tikz-feynman}
\usepackage{contour}
\usetikzlibrary{shapes,arrows,positioning,automata,backgrounds,calc,er,patterns}
\tikzfeynmanset{every blob={/tikz/fill=none, /tikz/inner sep=1pt}}
\makeatletter

\begin{document}
\preprint{INHA-NTG-06/2024} % NR: what does it mean?
\title{Heavy-light quark systems from the QCD instanton vacuum:
   $N_f=1$ light flavor case}  

\author{Ki-Hoon Hong}
\email{kihoon@inha.edu}
\affiliation{Department of Physics, Inha University, Incheon 22212,
  Republic of Korea} 

\author{Hyun-Chul Kim}
\email{hchkim@inha.ac.kr}
\affiliation{Department of Physics, Inha University, Incheon 22212,
  Republic of Korea} 
\affiliation{School of Physics, Korea Institute for Advanced Study
  (KIAS), Seoul 02455, Republic of Korea} 

\author{M.~M. Musakhanov}
\email{musakhanov@gmail.com}
\affiliation{Institute of Theoretical Physics, National University of Uzbekistan,
Tashkent 100174, Uzbekistan}

\author{N. Rakhimov}
\email{n.r.rakhimov@gmail.com}
\affiliation{CENuM, Korea University, Seoul, 02841, Republic of Korea} 

\begin{abstract}
We investigate heavy-light quark systems within the framework of the
QCD instanton vacuum, focusing on the $N_f = 1$ light flavor case. We
derive an effective heavy-light quark interaction from the low-energy
QCD partition function and construct a heavy-meson effective
Lagrangian. The physical residual mass of heavy mesons, $\Lambda$, is
determined by employing compositeness and normalization
conditions. We calculate the masses of $D$ and $B$ mesons and their
weak decay constants to the leading order and next-to-leading
order in the $1/m_Q$ expansion. The current results for $f_D$ and
$f_B$ are in good agreement with recent lattice QCD data and
Particle Data Group (PDG) average values. 
\end{abstract}
%\pacs{12.38.Lg,  12.39.Pn, 14.40.Pq}

\date{\today}
\maketitle

\section{Introduction}
A conventional heavy meson contains a heavy quark (antiquark) and a
light antiquark (quark). In the limit of infinitely heavy quark mass,
i.e., $m_Q\to \infty$, the spin of the heavy quark $\bm{S}_Q$ is
conserved. Consequently, the spin of the light quark is also
conserved: $\bm{S}_L=\bm{S}-\bm{S}_Q$, where $\bm{S}$ is the spin of
the heavy meson. This is known as heavy-quark spin
symmetry~\cite{Isgur:1989vq, Georgi:1990um, 
  Georgi:1991mr, Wise:1993wa}, which makes the spin of the light quark
a good quantum number. In this limit, a charm quark becomes
indistinguishable from a bottom quark, leading to heavy-quark
flavor symmetry. This implies that the infinitely heavy quark remains
as a mere static color source in a heavy-light quark system, so the
quark-gluon dynamics inside it is governed by the light quarks.  
Thus, heavy mesons are classified as
the members of the flavor SU(3) representations: $(D,\,D_s)\in
\bm{3}_0$ and $(D^*,\,D_s^*)\in \bm{3}_1$, where the subscripts denote
the corresponding spin of the heavy meson. Heavy mesons with a
light antiquark belong to the antitriplet ($\overline{\bm{3}}$). Since
light quarks dictate the dynamics inside heavy mesons, chiral
symmetry and its spontaneous breakdown come into crucial play in
describing the heavy mesons. Thus, as pointed out in
Refs.~\cite{Georgi:1991mr, Wise:1993wa}, both 
heavy-quark spin-flavor symmetry and chiral symmetry must be
considered when constructing an effective theory for the heavy-light
quark system.   

The instanton vacuum provides a theoretical framework for
realizing the spontaneous breakdown of chiral symmetry (SB$\chi$S) in
quantum chromodynamics (QCD)~\cite{Shuryak:1981ff, Diakonov:1983hh,
  Diakonov:1985eg} (see also the reviews~\cite{Schafer:1996wv,
  Diakonov:2002fq}). In Ref.~\cite{Diakonov:1983hh}, it was
quantitatively shown that the instanton medium, which consists of the
superposition of the instantons ($I$'s) and anti-instantons
($\bar{I}$'s), was stabilized at $R/\rho\approx 3$, where $R$ and
$\rho$ denote the average distance between $I$'s ($\bar{I}$'s) and the
average size of the instanton.   
$R$ and $\rho$ were determined to be $R\simeq 1\,\mathrm{fm}$ and  
$\rho\simeq 1/3\,\mathrm{fm}$~\cite{Shuryak:1981ff, 
  Shuryak:1982hk, Diakonov:1983hh,
  Diakonov:1985eg}. Both $R$ and $\rho$ are related to 
$\Lambda_{\mathrm{QCD}}$, allowing their determination once
$\Lambda_{\mathrm{QCD}}$ is given. Lattice QCD results support these
values~\cite{Chu:1994vi, deForcrand:1997esx, DeGrand:1997gu}. 
These $R$ and $\rho$ values well explain the quark and gluon
condensates, which characterize the QCD vacuum. The gluon condensate,
proportional to the instanton density $N/V\equiv 1/R^4$\cite{Diakonov:2002fq}, agrees with the phenomenological
estimate $\langle F_{\mu\nu}F_{\mu\nu} \rangle / 32\pi^2 \simeq(200\,
\mathrm{MeV})^4 > 0$\cite{Shifman:1979uw}. The quark condensate, the
order parameter for SB$\chi$S,
is determined as~\cite{Diakonov:1985eg}: 
\begin{align}
  \label{eq:1}
\langle \bar{\psi} \psi \rangle =
  -\frac1{R^2\rho}\sqrt{\frac{N_c}{6.62}} \simeq  - (253\,
  \mathrm{MeV})^{3} .
\end{align}
at the scale of $\mu=600$ MeV, which is approximately identified by
$\rho^{-1}\simeq 600$ MeV~\cite{Diakonov:2002fq, Kim:1995bq}. Here,
$N_c$ is the number of colors. The SB$\chi$S mechanism involves quark
propagation through the instanton medium. As quarks hop between random
$I$'s and $\bar{I}$'s, their helicity changes by different handedness
of $I$ and $\bar{I}$ zero modes. This delocalization process realizes
the spontaneous breakdown of chiral symmetry. This indicates that the
quark acquires the dynamical quark mass, $M(k)$, which is
momentum-dependent. $M(k)$ is related to the 
Fourier transform of the fermionic zero mode~\cite{Diakonov:1985eg,
  Diakonov:2002fq}. The value $M(0)\sim 350\,\mathrm{MeV}$ is
determined by the saddle-point equation. This framework also yielded
$f_\pi\simeq 93\,\mathrm{MeV}$~\cite{Diakonov:1985eg}. The effective
low-energy QCD partition function derived from the instanton vacuum
has been successfully applied to low-lying light hadrons and singly
heavy baryons in the limit of the infinitely heavy-quark
mass~\cite{Diakonov:1987ty, Musakhanov:2002xa, Nam:2007gf, Nam:2007fx, 
  Nam:2011yw, Son:2015bwa, Shim:2017wcq, Shim:2018rwv, 
  Christov:1995vm, Diakonov:1997sj, Yang:2016qdz, Kim:2021xpp}.  

Instanton effects on quarkonia were also studied and found to be
marginal. The instanton contribution to the heavy quark mass is
$\Delta M_Q\simeq 70\,\mathrm{MeV}$\cite{Callan:1978ye,
  Diakonov:1989un, Yakhshiev:2016keg}, which is small compared to the
charm quark mass. However, these effects allow for the 
values for parameters closed to physical ones in the heavy-quark
potential, such as the strong coupling constant and heavy-quark
mass. This approach has led to quantitative descriptions of charmonia
spectra and decay rates\cite{Yakhshiev:2016keg, Yakhshiev:2018juj,
  Yakhshiev:2021jkc, Hong:2022sht,  Musakhanov:2020hvk}. The
instanton vacuum thus provides a unified framework for investigating
both light-quark and heavy-quark systems, making it crucial to explore
effective interactions for heavy-light quark systems. Recently,
effective heavy-light quark interactions from the instanton vacuum
were sketched in Refs.~\cite{Chernyshev:1995gj,
  Musakhanov:2021gof}. In the present work, 
we aim to quantitatively complete the work of
Ref.~\cite{Musakhanov:2021gof} by constructing a heavy-meson effective
Lagrangian from the instanton vacuum. For simplicity, we focus on the
case of the light-quark flavor number $N_f=1$ and will present more
complicated and general cases elsewhere.   

This paper is organized as follows: 
in Section~\ref{sec:2}, we show how to derive the
heavy-light quark interaction from the instanton vacuum.
In Section~\ref{sec:3} we simplify this interaction and obtain
the four-quark interaction for $N_{f}=1$. In Section~\ref{sec:4}, we
derive the heavy meson residual mass $\Lambda$, a key parameter
predicted to be generated by nonperturbative corrections. We also
calculate the heavy meson masses, $M_D$ and $M_B$, and derive the weak
decay constants $f_D$ and $f_B$ to the $1/m_Q$ leading order and
next-to-leading order. The results are discussed in comparison with
other works. The final section summarizes our findings and presents an
outlook for future work.

%%%%%%%%%%%%%%%%%%%%%%%%%%%%%%%%%%%%%%%%%%%%
\section{Heavy-light quark systems
  from the instanton vacuum  \label{sec:2}}%  
%%%%%%%%%%%%%%%%%%%%%%%%%%%%%%%%%%%%%%%%%%%%
To derive the effective low-energy QCD partition function for the
light quark degrees of freedom, we need to compute the quark
determinant in the presence of the instantons and
antiinstantons, and then average it over the collective
coordinates of the instantons~\cite{Diakonov:1985eg,
  Musakhanov:1998wp, Diakonov:2002fq}. We will briefly recapitulate
the derivation of the partition function for the light quark systems
and extend it to the heavy-light quark systems. 
We start from the quark determinant derived from the instanton
vacuum~\cite{Diakonov:1985eg, Musakhanov:1998wp, Diakonov:2002fq,
  Musakhanov:2002xa, Goeke:2007bj}:
\begin{align}
\mathrm{Det}_N = &\int D\psi D\psi^\dagger \exp\left(\sum_f\int 
                   d^4x\,\psi^\dagger_f(i \Slash{\partial} + im_f)\psi_f\right)
            \prod_f \left\{\prod_+^{N_+} V_+[\psi^\dagger_f,\psi_f]
                   \prod_-^{N_-} V_- [\psi^\dagger_f,\psi_f]
                   \right\},
 \label{eq:2}                   
\end{align}
where $m_f$ denotes the current quark mass with flavor $f$. $N_+$ and
$N_-$ stand for the numbers of the instantons and antiinstantons,
respectively. $ V_\pm[\psi^\dagger_f,\psi_f]$ are the instanton-induced
$2N_f$ quark-quark interactions written as~\cite{Goeke:2007bj}:
\begin{align}
  \label{eq:3}
    V_\pm[\psi^\dagger_f,\psi_f] =& \int d^4x\left(\psi^\dagger(x)(i
     \rlap{/}{\partial})\Phi_\pm(x-z_\pm,U_\pm)\right)
     \int  d^4y \left(\Phi^\dagger_\pm
     (y-z_\pm,U_\pm)(i\rlap{/}{\partial} )\psi(y)\right).   
\end{align}
$\Phi_\pm$ is the fermionic zero mode obtained in the presence of the 
(anti)instanton field. We need to average the quark determinant over
the instanton zero modes $\xi_\pm$ that consist of its position, its 
size, and its orientation in color space.  The instanton medium
has low density, since its packing fraction is very small ($\pi^2
(\rho/R)^4\approx 1$). This diluteness of the instanton medium allows
us to perform the integration over the zero modes of the instantons
independently. Before we proceed to derive the effective low-energy
QCD partition function, we want to mention about the current quark
mass $m_f$. In the current work, we focus on the $N_f=1+1$ case, and
examine the heavy-quark spin-flavor symmetry and the effects of
$1/m_Q$ corrections. Thus, effects of flavor SU(3) symmetry breaking
will not be considered, so the chiral limit will be taken from now
on. 

After averaging the quark determinant over the zero
modes, we obtain the effective low-energy QCD partition 
function~\cite{Diakonov:1992bi,Diakonov:1995ea} as follows: 
\begin{align}
    \mathrm{Z}_{\mathrm{eff}} = & \int D\psi
    D\psi^\dagger\exp\left(\int d^4x\,\sum_f^{N_f}\psi^\dagger_f
          i\Slash{\partial}\psi_f\right)
          Y^{N_+}_{N_f} Y^{N_-}_{N_f},  
    \label{Z_light}
\end{align}
where $N_{f}$ denotes the number of light flavors,
$Y_{N_f}^{N_{\pm}}$ represent the instanton-induced $2N_{f}$
quark-quark interactions:
\begin{align}
    Y_{N_f}^{N_{\pm}} =& (i)^{N_f}
                      \left(\frac{4\pi^2\rho^2}{N_c}\right)^{N_f}
                      \int \frac{d^4 z}{V} \mathrm{det}_f i J_{\pm}(z)
\end{align}
with
\begin{align}
  \label{eq:5}
   J_{\pm}(z)_{fg} &= \int \frac{d^4 k d^4 l}{(2\pi)^8} \exp[-i(k-l)z]
                     F(k^2) \psi_f^\dagger (k) \frac12 (1\pm \gamma_5)
                     \psi_g(l) F(l^2).
\end{align}
Here, $F(k)$ designates the form factor related to the
zero-mode wave function in momentum space $\Phi_{\pm}(k;\xi_{\pm})$.
It is expressed in terms of the modified Bessel functions $I_i$ and
$K_i$ 
\begin{align}
  F(k)=-k\frac{d}{dk}\left[I_{0}\left(\frac{k\rho}{2}\right)K_{0}
  \left(\frac{k\rho}{2}\right)-I_1\left(\frac{k\rho}{2}\right)K_1
  \left(\frac{k\rho}{2}\right)\right].
\end{align}
The effective low-energy QCD partition function can further be
bosonized. For details, we refer to Refs.~\cite{Diakonov:1985eg,
  Diakonov:2002fq}. 

We now turn our attention to the heavy-light quark system.
Since the heavy quark mass is much larger than
$\Lambda_{\mathrm{QCD}}$, we consider the limit $m_Q\to\infty$. In
this limit, the heavy quark and antiquark fields are separate, as
$Q\bar{Q}$ pair creation does not occur, preventing overlap between
the heavy quark and antiquark functional spaces. Furthermore, due to
the infinite mass of the heavy quark, both the heavy quark and light
quark spins are conserved. Consequently, the low-lying heavy mesons
are classified as members belonging to either the flavor triplet
($\bm{3}$) or antitriplet ($\bar{\bm{3}}$) representations.

In the limit of $m_Q\to\infty$, the heavy quark becomes a mere
static color source. Thus, the heavy quark inside a heavy hadron is
considered to be almost on shell, so that its momentum can   
be decomposed as $p_{Q}^\mu=m_Qv^\mu+k^\mu$, where $v_\mu$ is the 
heavy quark velocity and $k_\mu$ denotes the residual momentum. Note
that the velocity of the heavy quark remains unchanged in the
presences of the interactions due to $k/m_Q\ll 1$. This leads to
the velocity superselection rule~\cite{Georgi:1990um}.
We introduce the covariant velocity-dependent projection operator
$P_\pm=(1\pm\Slash{v})/2$, which can be used to
project out particle and antiparticle states, $Q_\pm=P_\pm\,Q$. In
the rest frame, it becomes $(1+\gamma_0)/2$, selecting the particle
components from the Dirac spinor. We then redefine the heavy quark
field by extracting the dominant part of the heavy-quark momentum: 
\begin{align}
  Q(x) &=e^{-im_Qv\cdot x}(h(x)+\mathcal{H}(x)),\label{HQF(x)}
\end{align}
where $h(x)$ is the large component of $Q(x)$
\begin{align}
  h(x)&=e^{im_Qv\cdot x}Q_+(x),
    \label{def:h}
\end{align}
and $H(x)$ is the small component given in the order of $1/m_Q$  
\begin{align}
  \mathcal{H}(x)=e^{im_Qv\cdot x}Q_-(x).
\end{align}
The $h$ field contributes in the leading order, while the ${\cal H}$
field produces effects suppressed by factors of
$1/m_Q$. $\mathcal{H}$ can be effectively eliminated by integrating
out the field using the Dirac equation. With the inverse heavy-quark
mass $1/m_Q$ as an expansion parameter, we arrive at the heavy-quark 
effective Lagrangian~\footnote{Since the instantons are defined in
  Euclidean space, we also express the effective Lagrangian in
  Euclidean space.}: 
\begin{align}
    {\cal L}_{\mathrm{eff}} = -h^\dagger i(v\cdot D)h -\frac{i}{2m_Q}
  h^\dagger
  \left(D_{\perp}^2-\frac{g}{2}\sigma_{\mu\nu}G^{\mu\nu}\right)h, 
    \label{HQET_Lag}
\end{align}
where $D_\perp$ is a component of the covariant derivative transversal
to velocity $v$, defined as $D_{\perp\mu}=D_\mu-v_\mu v\cdot D$. 
The lowest-order Lagrangian, the first term of Eq.~\eqref{HQET_Lag},
reflects heavy-quark spin-flavor symmetry. 

The heavy quark propagator in the instanton ensemble can
be written in terms of the propagators in the single instanton
background. Due to the small packing fraction, $\rho^4N/(VN_c)\approx
0.004$, we can express the heavy quark propagator as
$w=\left(i\partial_4-i\sum_\pm
  A_{\pm,4}\right)^{-1}$~\cite{Diakonov:1989un,Pobylitsa:1989uq} in 
the rest frame $v_\mu=(\vec 0,1)$. The inverse of the propagator is
written as    
\begin{equation}
    \bar w^{-1} = i\partial_4-\sum_\pm N_\pm \left\langle
      i\partial_4\left(w_\pm-(i\partial_4)^{-1}\right)i \partial_4
    \right\rangle,   
    \label{Pob-eq}
\end{equation}
where $\langle\hat{\cal O}\rangle = \int d\xi_\pm\hat{\cal O}$ denotes
averaging over the single-instanton fields.
$w_\pm=\left(i\partial_4-iA_{\pm,4}\right)^{-1}$ represents the heavy
quark propagator in the single-instanton (antiinstanton) field.
The heavy-quark propagator can be expressed as the Wilson line in large
Euclidean time, allowing us to calculate the instanton contribution to
the heavy quark mass $\Delta M_Q\simeq70$ MeV. This Wilson line yields
the instanton-induced heavy-quark potential\cite{Diakonov:1989un}.
Although the instanton effects on heavy quarkonium spectra are
marginal~\cite{Yakhshiev:2016keg, Yakhshiev:2018juj,
Musakhanov:2020hvk, Hong:2022sht},
they enable one to use almost physical values values for parameters
such as the strong coupling constant and the charm quark mass. Note
that these parameters are often treated as free variables to better
describe quarkonium physics phenomenologically
\cite{Yakhshiev:2016keg, Yakhshiev:2018juj, Hong:2022sht}. 

Combining Eqs.~\eqref{Z_light} and \eqref{HQET_Lag}, we can derive
the instanton-induced heavy-light quark interaction. This interaction
depends on the total flavor number $N_f = N_{lf} + N_{hf}$, where
$N_{lf}$ and $N_{hf}$ represent the numbers of light and heavy
flavors, respectively. In the limit of $m_Q\to \infty$, the heavy
quark becomes independent of its flavor, so we can express the total
flavor number as $N_{f}=N_{lf}+1$. In the following, we will derive
the $N_f$ instanton-induced heavy-light quark interaction.  

We are now in a position to discuss both heavy and light quarks in the 
$N_\pm$ instanton ensemble. The corresponding fermionic determinant
can be written as a product of light and heavy ones. We will replace
the light-quark determinant with an approximate one. The effective
low-energy QCD partition function can then be written as  
\begin{align}
{\cal Z}_{\mathrm{eff}}=&\left(\prod_\pm^{N_\pm} \int
 d\xi_\pm\right)\,\int D\psi D\psi^\dagger
  \prod_f^{N_{lf}}\exp\left(\int d^4x\,\psi^\dagger_f
 \,i\Slash{\partial}\,\psi_f\right)  \prod_\pm^{N_\pm}
  V_\pm[\psi^\dagger_f,\psi_f]\cr 
  &\times \int Dh Dh^\dagger
    \exp\left(\int d^4x\, h^\dagger
    \left(i\partial_4-i\sum_\pm A_{\pm,4}\right)h\right),
    \label{Z_Qq}
\end{align}
where $\left(\prod_\pm^{N_\pm} \int d\xi_\pm\right)$ represents
the average over the instanton collective coordinates. While one can
calculate the heavy-quark propagator from Eq.~\eqref{Z_Qq}, 
averaging over the instanton collective coordinates prevents immediate
integration over light quark degrees of freedom. To overcome this, we
introduce the identity operator 
\begin{align}
    \mathbf{1} = \left(\prod_\pm^{N_\pm} \left \langle \prod_f 
V_\pm[\psi^\dagger_f,\psi_f] \right\rangle\right) 
\left(\prod_\pm^{N_\pm}\left\langle 
\prod_fV_\pm[\psi^\dagger_f,\psi_f] \right\rangle\right)^{-1}.
\label{eq:14}
\end{align}
We can group the first part of Eq.~\eqref{eq:14} together with the
kinetic term for the light quarks. We can then add the light quark
source fields, following the approach in Refs.~\cite{Kim:2004hd,
  Goeke:2007nc}. This procedure allows us to integrate over the
collective coordinates for the heavy-quark part
separately: 
\begin{align}
    w[\psi^\dagger,\psi]=&\left(\prod_\pm^{N_\pm}\left\langle 
\prod_f^{N_{lf}}V_\pm[\psi^\dagger_f,\psi_f] \right\rangle\right)^{-1}
\left(\prod_\pm^{N_\pm}\int d\xi_\pm\right) 
\prod_f^{N_{lf}}V_\pm[\psi^\dagger_f,\psi_f]
\left(i\partial_4-i\sum_\pm A_{\pm,4}\right)^{-1}.
    \label{wpsi}
\end{align}
We extend the method in Refs.~\cite{Diakonov:1989un, Pobylitsa:1989uq}
to the current case. Using the packing faction $\rho^4N/(VN_c)\ll 1$
as a perturbative expansion parameter, we obtain an iterative solution up
to linear order. This leads to an averaged inverse 
functional $w^{-1}[\psi^\dagger,\psi]$ in terms of a 
single-instanton average:  
\begin{align}
     \bar w^{-1}[\psi^\dagger,\psi]
    = i\partial_4-\sum_\pm N_\pm& \left\langle\prod_f^{N_f}
V_\pm[\psi_f^\dagger,\psi_f]\right\rangle^{-1}
\Delta_{H,\pm}[\psi^\dagger,\psi], 
    \label{eq:16}
\end{align}
where 
\begin{eqnarray}
    \Delta_{H,\pm}[\psi^\dagger,\psi] &=& \left\langle
    \prod_f^{N_f}V_\pm[\psi_f^\dagger,\psi_f] % light part
    i\partial_4\left(w_\pm-(i\partial_4)^{-1}\right)i\partial_4 % heavy part
    \right\rangle.
    \label{Delta-H}
\end{eqnarray}
To derive the heavy-light quark interaction, we first combine
Eq.~\eqref{eq:16} with the light quark part to obtain the heavy-quark
propagator averaged over (anti)instanton collective coordinates:
\begin{align}
    \bar w=&\frac{1}{{\cal Z}[0]}\int D\psi^\dagger D\psi
             \prod_\pm\int d\lambda_\pm \exp\int d^4x\left\{
             \psi_f^\dagger\, i\Slash{\partial}\, \psi_f+\lambda_\pm
             \left\langle\prod_f^{N_f}V_\pm[\psi^\dagger,\psi]
             \right\rangle \right\}\cr  
    &\times\left(i\partial_4-\sum_\pm\lambda_\pm
      \Delta_{H,\pm}[\psi^\dagger,\psi]\right)^{-1}  
    \label{HQ_prop}
\end{align}
Here, the Lagrange multiplier $\lambda_\pm$ has to be integrated in
the saddle point approximation. The function inside the exponent is
responsible for the saddle point equation. Thus, we can safely say
that the saddle point value $\lambda_\pm=\lambda_{\rm
  s.p.}$ is defined in the light quark sector. According to
Refs.~\cite{Diakonov:1992bi,Diakonov:1995ea} it defines the light
quark dynamical mass $M(k) = \lambda_{\rm s.p.}\left[2\pi\rho
  F(k)\right]^2/N_c$. Next, we can rewrite the heavy-quark propagator 
in terms of the heavy-quark fields $h$ and $h^\dagger$ in the
form of the functional integration: 
\begin{align}
    \bar w=&\frac{1}{{\cal Z}[0]}\prod_\pm\int d\lambda_\pm \int
             D\psi^\dagger D\psi \exp\int d^4x\left\{\psi_f^\dagger
             \,i\Slash{ \partial}\,\psi_f+\lambda_\pm
             \left\langle\prod_f^{N_f}V_\pm[\psi^\dagger,\psi]
             \right\rangle\right\} \cr  
    &\times\int Dh^\dagger Dh\,h h^\dagger\exp\int d^4x\left\{
      h^\dagger\left(i\partial_4-\sum_\pm\lambda_\pm
      \Delta_{H,\pm}\right)h\right\}.  
    \label{HQ_prop_h}
\end{align}
From the last term of Eq.~\eqref{HQ_prop_h}, we can identify the
heavy-light quark interaction. Using definitions
(\ref{Delta-H}) and (\ref{eq:3}) one can express the interaction as   
\begin{align}
    &S_{\rm int} 
    = \sum_\pm \int
      d^4z_\pm\,dU_\pm\,\frac{N_\pm}{V}\prod_{f=1}^{N_{lf}}\left\{\frac{N_cV}{N_\pm} 
\int \frac{d^4k_f d^4l_f}{(2\pi)^8}e^{i(k_f-l_f)\cdot z_\pm}
\sqrt{M(k_f)M(l_f)}\right.\cr 
    &\left.\times \psi^\dagger_{\alpha_fi_f}(k_f)
\left(\gamma_\mu\gamma_\nu\gamma_\pm\right)^{i_f}_{j_f}U^{\alpha_f}_{\pm,\gamma_f}
\left(\tau^\mp_\mu\tau^\pm_\nu\right)^{\gamma_f}_{\delta_f}
U^{\dagger \delta_f}_{\beta_f}\psi^{\beta_fj_f}(l_f)\right\}\cr  
    &\times\int d^3x\,dt\,dt'\,\, h^\dagger_{\alpha_3}(\vec x, t')
U^{\alpha_3}_{\pm,\gamma_3}\langle
      t'|i\partial_4(w_\pm-(i\partial_4)^{-1})
i\partial_4|t\rangle^{\gamma_3}_{\delta_3} U^{\dagger
      \delta_3}_{\pm,\beta_3}
h^{\beta_3}(\vec x, t).\label{SQq_Nf}
\end{align}
To obtain the effective interaction with $N_{lf}$ fixed, one must
perform color integration and Fierz reordering. In the next 
section, we will discuss the instanton-induced $N_f=1+1$ effective
heavy-light interaction and its applications. 

%%%%%%%%%%%%%%%%    N_{lf} = 1    %%%%%%%%%%%%%%%% 
\section{$N_{f}=1+1$ heavy-light interaction}\label{sec:3}
%%%%%%%%%%%%%%%%%%%%%%%%%%%%%%%%%%%%%%%%%%%%%%%
The heavy quark fields can be represented in momentum
space, resulting in integration over each quark momentum followed by a 
$\delta$-function defining energy-momentum conservation.
In this case, the color-orientation integral can be done
straightforwardly as shown in Appendix~\ref{app:1}. 
To obtain color and spin contracted
terms, we will use the notation 
$\psi^\dagger_{ia}\psi^{ia}=\left(\psi^\dagger\psi\right)$, 
where $i$ and $a$ represent spin and color the indices, respectively.
Then we carry out Fierz reordering. Furthermore, following
Ref.~\cite{Diakonov:1989un}, in $m_Q\to\infty$ limit, we can 
derive the instanton effect on the heavy-quark mass 
\begin{align}
    \frac{1}{N_c}\int d^3xdt dt'\,\, e^{i\vec{p}\cdot\vec{x}}\tr_c
  \langle
  -t'|i\partial_4(w_\pm-(i\partial_4)^{-1})i\partial_4|t
  \rangle\,\,\to\,\,   -\frac{V}{ N}\Delta M_Q.
    \label{T(p->0)}
\end{align}
Finally, in the large $N_c$ limit, we derive the instanton induced
$N_f=1+1$ effective heavy-light quark interaction given by 
\begin{align}
    S_{\rm int}&=-\int\frac{d^4k_1d^4k_2}{(2\pi)^8}
    \frac{d^4p_1d^4p_2}{(2\pi)^8}
    (2\pi)^4\delta^{(4)} \left(k_1-k_2+p_1-p_2\right)
    F(k_1)F(k_2)M_q(0)\frac{\Delta    M_Q}{N/V}\cr     
    &\times\left[\left(\psi^\dagger(k_1)\psi(k_2)\right)
       \left(Q_+^\dagger(p_1)        Q_+(p_2)\right) +
       \frac{1}{8}\sum_i\left(\psi^\dagger(k_1)\Gamma_i 
       Q_+(p_1)\right)\left(Q_+^\dagger(p_2)\Gamma_i
       \psi(k_2)\right)\right],
\label{S_int} 
\end{align}
where $\Gamma_i \in\left\{
  \mathbf{1},~\gamma_5, ~\gamma_\mu,~ i\gamma_\mu\gamma_5,
  ~\sigma_{\mu\nu}/ \sqrt{2}\right\}$ denote the Dirac matrices.  
This interaction can be understood as follows: we began with heavy and
light quarks in the background of the instanton medium. The dilute nature of
this instanton medium allowed us to rewrite the QCD partition function
as the partition function of the $N_\pm$-instanton configuration
multiplied by the instanton ensemble averaged action for heavy and
light quarks. By performing a perturbative expansion with respect to
the instanton packing fraction enable us to simplify the ensemble
average to the average over a single (anti-)instanton. This leads to
an effective four-quark interaction between the heavy and light quarks,
with a form factor $F(k)$ attached to each light quark leg. This
interaction is characterized by a coupling $g^2 \sim M_q(0)\Delta M_Q
R^4$, where $M(0)$ is the dynamical light quark mass, $\Delta M_Q$ is
the instanton contribution to the heavy quark mass, and $R$ is the
average separation between instantons.  

We now express the second term in the bracket of Eq.~\eqref{S_int} in
coordinate space, where the form factor $F$ can be written as a function
of the derivative 
\begin{align}
S_{\rm int}&=-g^2\int d^4x \left(Q_+^\dagger(x)\Gamma_i
F(\partial)\psi(x)\right)\left(F(\partial)
\psi^\dagger(x)\Gamma_i   Q_+(x)\right),  
\qquad g^2=M_q(0)\Delta M_Q R^4/8,
\label{eq:quark-quark}
\end{align}
where $g^2$ is given in the large $N_c$ limit.
To further develop this model, we can employ the bosonization. This
involves replacing the heavy-light bilinears
$\left(h^\dagger\Gamma_i F(\partial)\psi\right)$ with corresponding
auxiliary boson fields $\Phi_i$. This transformation yields the
effective Lagrangian describing the interaction between heavy
mesons and heavy-light quarks 
\begin{align}
    \mathcal{L}_{\mathrm{eff}} = -\psi^\dagger\left(i\Slash{\partial} + i
  M_q(\partial) \right)\psi -h^\dagger\, i(v\cdot\partial)h + m_\Phi^2
  \Phi_i^\dagger\Phi_i -G\Phi_i^\dagger F(\partial) \psi^\dagger
  \Gamma_i Q_+ - G Q_+^\dagger \Gamma_i F(\partial)\psi\Phi_i,
    \label{L_eff}
\end{align}
where $G=gm_\Phi$ and $m_\Phi$ stands for the mass of $\Phi$. The
attractive force between the heavy quark and the light antiquark is
ensured by $G>0$, guaranteeing the production of the heavy-light
composite particle.  
Next, a renormalization process must be performed to identify the
auxiliary fields as corresponding physical meson fields. This crucial 
step will be discussed in detail in the following section. 
%%%%%%%%%%%%%%%%%%%%%%%%%%%%%%%%%%%%%%%%%%%%%%%%%%%%%%
\section{Equivalent effective Lagrangian\label{sec:4}}
%%%%%%%%%%%%%%%%%%%%%%%%%%%%%%%%%%%%%%%%%%%%%%%%%%%%%%
It is important to note that the auxiliary field $\Phi$ in 
Eq.~\eqref{L_eff} does not yet represent the physical heavy meson
field, as the bosonization process does not inherently impose any 
physical constraints. Therefore, we need to ensure that this 
field is properly renormalized such that it acquires a physical
mass. Once the compositeness condition $Z_\Phi=0$ is considered, where
$Z_\Phi$ denotes the wavefunction renormalization constant, we will
show that Eq.~\eqref{eq:quark-quark} is equivalent to the following
Lagrangian~\cite{Shizuya:1979bv} 
\begin{align}
 \mathcal{L}_{\mathrm{eff}}'= - \psi^\dagger\left(i\Slash{\partial} +
  iM_q(\partial) \right)\psi -h^\dagger\,i(v\cdot\partial)h
  + \partial_\mu
  \Phi^\dagger_i\partial_\mu\Phi_i+m_\Phi^2\Phi_i^\dagger\Phi_i 
  -G\Phi_i^\dagger F(\partial)\psi^\dagger\Gamma_iQ_+ - GQ_+^\dagger
  \Gamma_i F(\partial)\psi\Phi_i. 
    \label{eq:L_kin}
\end{align}
The $\Phi$ field and its mass in Eq.~\eqref{eq:L_kin} are required to
be renormalized. They are related to the renormalized ones as follows  
\begin{align}
&\Phi = Z_\Phi^{1/2}\Phi_R, \;\;\; Z_\Phi m_{\Phi}^2=m_{\Phi
  R}^2+\delta m_\Phi^2. 
\end{align}

We will briefly recapitulate how the compositeness condition can be
proved~\cite{Cheng1998,Lurie:1964}. In the limit $m_Q\to \infty$, the
kinetic term and mass term for $\Phi$ can be expressed to zeroth-order
of $1/m_Q$ as    
\begin{align}
\partial_\mu\Phi^\dagger\partial_\mu \Phi + m_\Phi^2\Phi^\dagger \Phi
  = \phi_v^\dagger(-2v\cdot\partial-2\Lambda) \phi_v + \mathcal{O}
  \left(\frac{1}{m_Q}\right),
\label{eq:kinetic_heavymeson} 
\end{align}
where $\phi_v$ is related to $\Phi^\pm=\frac{1}{\sqrt{m_Q}}e^{\mp
  m_Qv\cdot x}\phi_v$ with $\Phi^+=\Phi$ and $\Phi^-=\Phi^\dagger$ in
Euclidean space. The heavy meson masses are defined as
$m_\Phi=m_Q+\Lambda$ in the infinitely heavy quark mass limit. The
term $\Lambda$ is known as the residual mass of the heavy meson. It is 
related to the heavy meson momentum $P_\Phi=m_\Phi
v=m_Qv+p_\Phi=m_Qv+\Lambda v$, which satisfies energy-momentum 
conservation: $P_\Phi=p_Q+p_q=m_Qv+k+p_q$, where $p_Q$ 
denotes a heavy quark momentum, and $p_q$ designates a light quark 
momentum. Consequently, the residual momentum for the heavy meson 
$p_\Phi=\Lambda v$ satisfies the 4-momentum conservation
$p_\Phi=p_q+k$.   

\begin{figure}[ht]
    \centering
    \subfigure[]{\includegraphics[width=5.8cm]{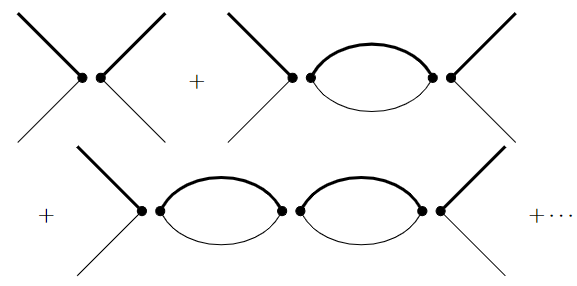}
\label{fig:quark-quark}}   
    \subfigure[]{\includegraphics[width=8cm]{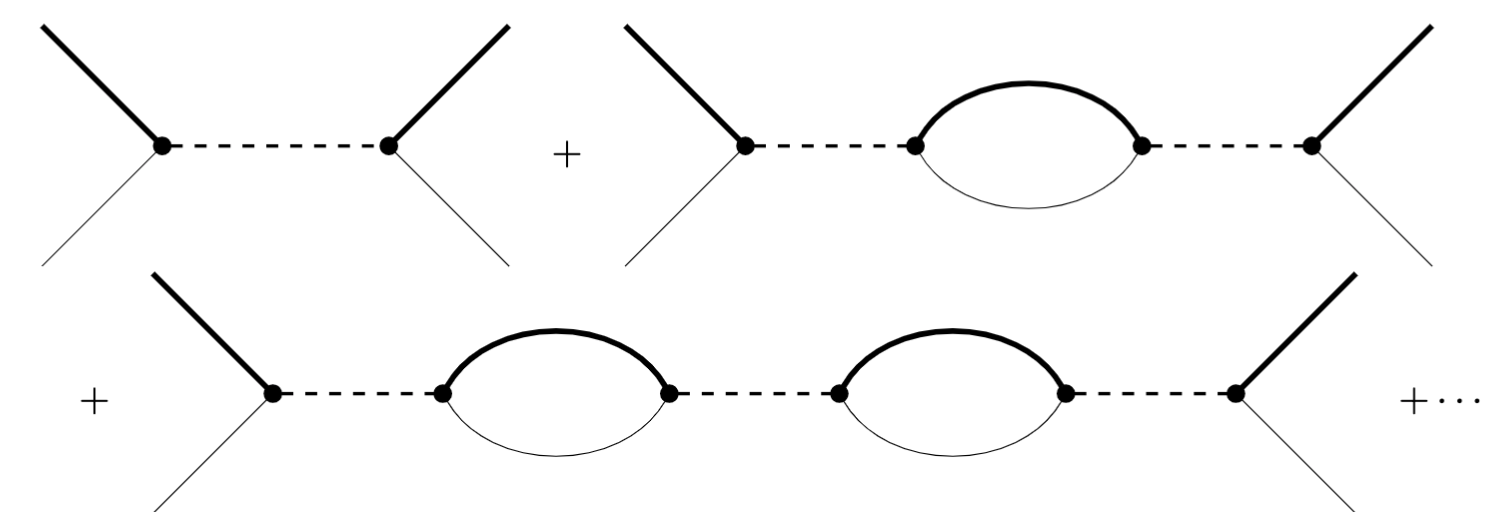}
\label{fig:meson-quark}}
\caption{In the left panel (a), the infinite sum for the quark-quark
  correlation function is drawn, whereas in the right panel (b) the
  meson-quark correlation function is depicted.}
\end{figure}

Then, we can rewrite the Lagrangian of Eq.~\eqref{eq:L_kin} as 
\begin{align}
    \mathcal{L}=-\psi^\dagger\left( i\Slash{\partial} +
  iM_q(\partial) \right)\psi -h^\dagger\,i(v\cdot\partial)h
  -\phi_{vi}^\dagger (2v\cdot\partial+2\Lambda) \phi_{vi} -
  G_0\phi_{vi}^\dagger F(\partial)\psi^\dagger\Gamma_ih - G_0h^\dagger
  \Gamma_i F(\partial)\psi\phi_{vi},
\label{eq:L_kin2} 
\end{align}
where $G_0$ is a rescaled coupling constant, which must be determined
by the renormalization.
For this Lagrangian to be the correct form of the effective
Lagrangian, correlation functions derived from
Eq.\eqref{eq:quark-quark} and Eq.\eqref{eq:L_kin2} must be physically
identical. We can compute the hevay-light quark correlation functions
using Eq.~\eqref{eq:quark-quark}.  
The momentum distribution of the quark inside a heavy meson is
governed by the quark form factor $F(p)$. Since the 
average of $gF(p)$ in momentum space, which can be regarded as an
effective coupling for the heavy-light quark vertex, is close to zero,
i.e., $\langle g F(p_q)\rangle\approx 0$, the infinite quark-loop and
meson-quark sums converge as shown in Figs.~\ref{fig:quark-quark} and
\ref{fig:meson-quark}, respectively. Similar loop calculations can be
found in Refs.~\cite{Lurie:1964,Cheng1998}. 
The corresponding expressions for the correlation functions are given by
\begin{align}
    \mathcal{A}^a_{Qq} &= \frac{-F(p_q)F(p_q')}{(v\cdot
  p_{\Phi}-i\Lambda_r) \Pi'(i\Lambda_r)+\Pi^r(i\Lambda_r)},\cr
    \mathcal{A}^b_{Qq} &= \frac{iG_0^2F(p_q)F(p_q')}{2(v\cdot
          p_\Phi-i\Lambda_r) - iG_0^2\Pi^r(i\Lambda_r)}, 
\end{align}
which determines the coupling constant
$G_0^2=2i/\Pi'(i\Lambda_r)$. Therefore, we obtain the compositeness 
condition~\cite{Shizuya:1979bv,Cheng1998} as follows 
\begin{align}
    Z_\Phi=\left(1+\frac{iG_0^2}{2}\Pi'(i\Lambda_r)\right)=0,
\label{eq:31}
\end{align}
where $\Lambda_r$ is a physical residual mass.
The fermion loop $\Pi$ is given as 
\begin{align}
    \Pi(v\cdot p_\Phi)=\int\frac{d^4p_q}{(2\pi)^4}F^2(p_q) \tr_D
  \left[ \Gamma_P\frac{(1+\Slash{v})}{2v\cdot(p-p_q) + i\epsilon}
  \Gamma_P \frac{-\left(\Slash{p}_q-iM(p_q)\right)}{p_q^2+M^2(p_q)}
  \right].
\label{eq:self_energy} 
\end{align}
$p=p_q+k$ corresponds to the definition of the heavy meson's residual 
momentum $p_\Phi$. Equation~\eqref{eq:31} leads to the equivalence
between the Lagrangians given in Eq.~\eqref{L_eff} and
Eq.~\eqref{eq:L_kin2}. The coupling constant $G_0$ can be represented
as $G_0=g\mathcal{N}G_1$, where $\mathcal{N}$ is a normalization
constant of heavy-meson state that will be introduced in the next
section, and $G_1$ only stands for carrying the dimension of $\rm
GeV^{1/2}$. 

\section{Normalization and residual mass of heavy meson}
\begin{figure}[htp]
    \centering
    \subfigure[]{
    \includegraphics[width=6cm]{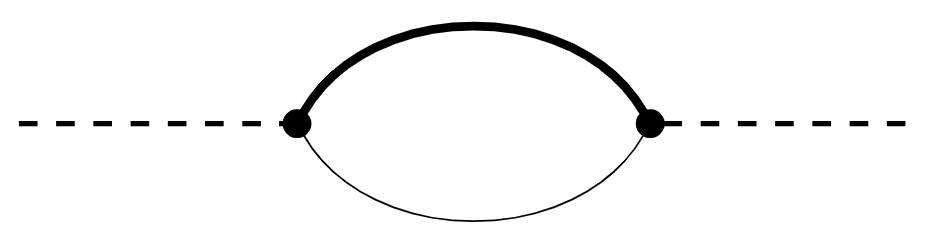}}
      \subfigure[]{\includegraphics[width=4cm]{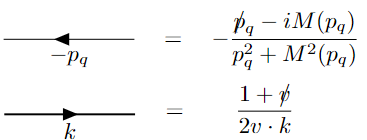}}
      \caption{(a) The diagrammatic form of the normalization for heavy
      meson states~\cite{Cheng:1997au,Hwang:2005uk} (b) The diagrammatic representations of the propagator of light and heavy quarks are $S_q(p_q)$ and $S_Q(k)$, respectively. } 
    \label{fig:2}
\end{figure}

We are now in a position to compute the renormalized residual mass of
the heavy meson. To obtain the physical mass, the auxiliary field
$\phi_v$ satisfies the standard normalization condition. The standard
relativistic normalization is given as 
\begin{align}
    \langle \Phi(\vec{p}')|\Phi(\vec{p})\rangle=(2\pi)^32p_4\delta^{3}(\vec{p}'-\vec{p}).
\end{align}
In $m_Q\rightarrow \infty$ limit, the heavy meson states can be
expressed as $|\phi(v)\rangle=m_{\Phi}^{-1/2}|\Phi(\vec{p})\rangle$,
then the nonrelativistic or heavy-quark mass independent normalization
can be represented by~\cite{Neubert:1993mb,Cheng1998} 
\begin{align}
    \langle\phi(v')|\phi(v)\rangle = (2\pi)^32p_4/m_\Phi \delta^{3}
  (\Lambda\vec{v}' -\Lambda\vec{v})=(2\pi)^3 2v_4 \delta^3
  (\Lambda\vec{v}' -\Lambda\vec{v}).
\label{normalization_cond}
\end{align}
To confirm Eq.~\eqref{normalization_cond}, let us define the wave
packet of the heavy meson in momentum space. We find its form from the
equation of motion from the effective Lagrangian~\eqref{L_eff}, where
the heavy-quark field is expressed as 
$\Phi_{i}\propto gF(\partial) \psi^\dagger\Gamma_iQ_+$. For
convenience, we introduce the factors $\sqrt{2v_4}\mathcal{N}$, where
$\mathcal{N}$ is a normalization constant. Then we can construct the
wave packet $\Phi_v$ in momentum space as follows 
\begin{align}
    \Phi_{vi}(p_\Phi)=\langle p_{\Phi4}|\phi_i(v)\rangle = \sqrt{2v_4}
  ig\mathcal{N}\int \frac{d^4p_qd^4k}{(2\pi)^8}(2\pi)^4
  \delta^{(4)}(p_\Phi-k-p_q)F(p_q)\psi^\dagger(p_q)\Gamma_i
  h(k),
\label{nonlocal_op} 
\end{align}
where the index $i$ denotes a type of bilinears, i.e., scalar, vector, 
pseudoscalar, and so on. The one-particle state for a heavy meson is
defined as $|\phi(v)\rangle\equiv\phi_v|0\rangle$. The form
factor $F(p_q)$ encodes the quark-instanton dynamics in the
light-quark sector, which also stabilizes the wave packet.  
A diagrammatic form of the normalization is shown in
Fig.~\ref{fig:2}. 

\begin{figure}[htp]
    \centering
    \includegraphics[width=10cm]{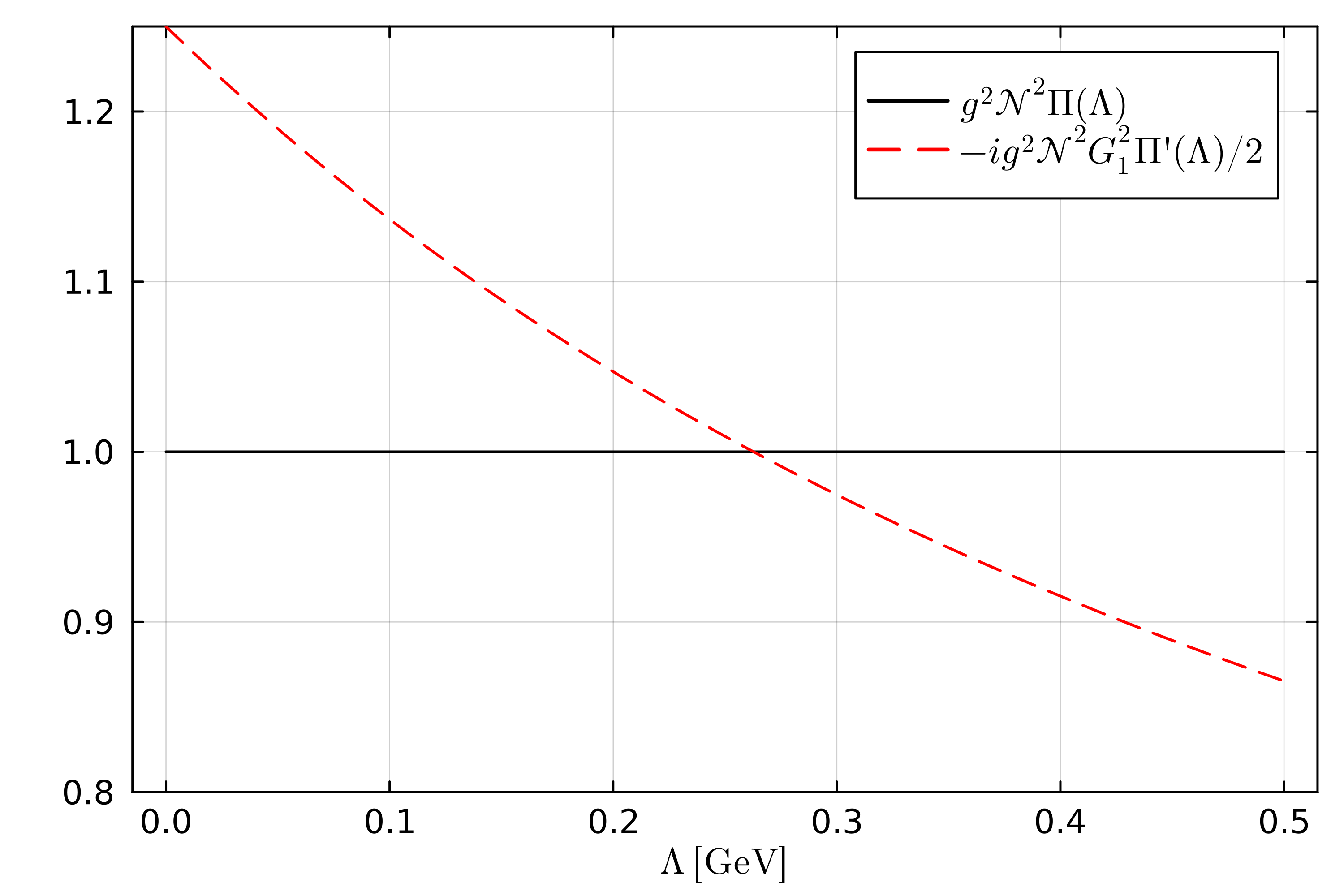}
    \caption{The black solid line and red dashed curve draw the
      normalization condition and the renormalization condition with
      the rescaled function $\Pi(\Lambda)$ by the normalization
      constant $\mathcal{N}(\Lambda)$, respectively. The intersection 
      point yields the physical residual mass, $\Lambda=0.2628\,\rm
      GeV$.}  
    \label{fig:residual_mass}
\end{figure}

It is straightforward to show that Eq.~\eqref{nonlocal_op} satisfies
the nonrelativistic normalization condition~\cite{Georgi:1991mr}
\begin{align}
\langle\phi_i(v')|\phi_j(v)\rangle &=\int\frac{dp_{\Phi4}}{2\pi}
   \tr_D\Phi_{vj}(p_\Phi)\Phi_{v'i}^\dagger(p'_\Phi)\cr  
&=2v_4g^2\mathcal{N}^2(2\pi)^3\delta^3(\Lambda v'-\Lambda v)
\int \frac{d^4p_q}{(2\pi)^4}F^2(p_q)\tr_D[S_q(p_q)\Gamma_i
S_Q(p_\Phi-p_q)\Gamma_j]\cr
&=2v_4g^2\mathcal{N}^2\Pi(i\Lambda)(2\pi)^3
\delta^3(\Lambda v'-\Lambda v)\delta_{ij}\cr
&=2v_4(2\pi)^3\delta^3(\Lambda v'-\Lambda v)\delta_{ij},
\end{align}
where we have replaced $\Lambda_r\rightarrow \Lambda$ for simplicity,
and the functions $S_q$ and $S_Q$ denote the light and heavy quark
propagators given in Eq.~\eqref{eq:self_energy}, as 
shown in Fig.~\ref{fig:2}. The normalization yields the normalization
condition, $g^2\mathcal{N}^2\Pi(i\Lambda)=1$. The 
physical residual mass $\Lambda$ is determined by the intersection of
the compositeness and normalization conditions, as shown in
Fig.~\ref{fig:residual_mass}. The result for $\Lambda$ is
$\Lambda\approx262.8\,\rm MeV$.   

$\Lambda$ is typically considered to be on the order of $\Lambda \sim
\Lambda_{\mathrm{QCD}}$~\cite{Falk:1992fm}, approximately $200\sim
300,$MeV, as it arises from heavy-light quark interactions in QCD. The
current approach, based on the instanton vacuum, explicitly
demonstrates this origin of $\Lambda$. This suggests that the residual
mass quantifies the contribution of heavy-light quark interactions to
a heavy meson mass, $m_\Phi$. Instanton effects are known to generate
dynamical quarks. We can decompose $m_\Phi$ as $m_\Phi=m_Q+\Delta
M_Q+M_q+\Lambda$, where $m_Q$ represents the charm (bottom) quark
mass, $\Delta M_Q$ denotes instanton effects on the heavy-quark mass,
and $M_q$ signifies the dynamical quark mass emerging from
SB$\chi$S. The final term, $\Lambda$, the physical residual mass, is
evaluated through the heavy-light quark mass. Consequently, we
determine the $D$ and $B$ meson masses to be $m_D=1.942,{\rm GeV},
(m_c=1.27,{\rm GeV})$~\cite{ParticleDataGroup:2024cfk} and
$m_B=5.322,{\rm  GeV},(m_b({\rm 1S\   scheme})={4.65,\rm
  GeV})$~\cite{ParticleDataGroup:2024cfk}, 
respectively. These values approximate the average masses of
pseudoscalar ($D,\, B$) and vector ($D^*,\, B^*$) mesons. In the
$m_Q\to\infty$ limit, the masses of the pseudoscalar and vector heavy
mesons are degenerate because of heavy quark-spin symmetry. This
degeneracy is shifted by the $1/m_Q$ correction given in Eq.~\eqref{HQET_Lag}. Its mesonic matrix element can be expressed as
\begin{align}
    \frac{1}{4m_Q}\langle \phi_{\vec{J}}|h^\dagger\sigma_{\mu\nu}hG_{\mu\nu}|\phi_{\vec{J}}\rangle&=\frac{1}{4m_Q}\langle h;\vec{S}_Q|h^\dagger\sigma_{\mu\nu}h|h;\vec{S}_Q\rangle\langle q;\vec{S}_q|G_{\mu\nu}|q;\vec{S}_q\rangle\cr
    &=\frac{\lambda_2}{m_Q}\langle 2\vec{S}_Q\cdot\vec{S}_q\rangle,
\end{align}
where the states $|\phi_{\vec{J}}\rangle\equiv|\phi(v);\vec{J};\vec{S}_Q;\vec{S}_q\rangle$ with the total spin $\vec{J}=\vec{S}_Q+\vec{S}_q$, and $|h(q);\vec{S}_Q(q)\rangle$ stands for each quark and its spin state. The eigenvalue of the heavy-light spin product is given as 
\begin{align}
    s_{Qq}&\equiv\langle 2\vec{S}_Q\cdot\vec{S}_q\rangle=J(J+1)-S_Q(S_Q+1)-S_q(S_q+1)\cr
    &=\left\{\begin{array}{cc}
    -3/2&\text{pseudoscalar}\\
    1/2&\text{vector}
    \end{array}
    \right..
\end{align}
To obtain $\lambda_2$, we have to compute the mesonic matrix element of the gluon operators from the instanton vacuum~\cite{Diakonov:1995qy}, which will be considered in future work.

\section{Weak decay constants of $D$ and $B$ mesons}
\begin{figure}[ht!]
    \centering
    \includegraphics[width=0.3\linewidth]{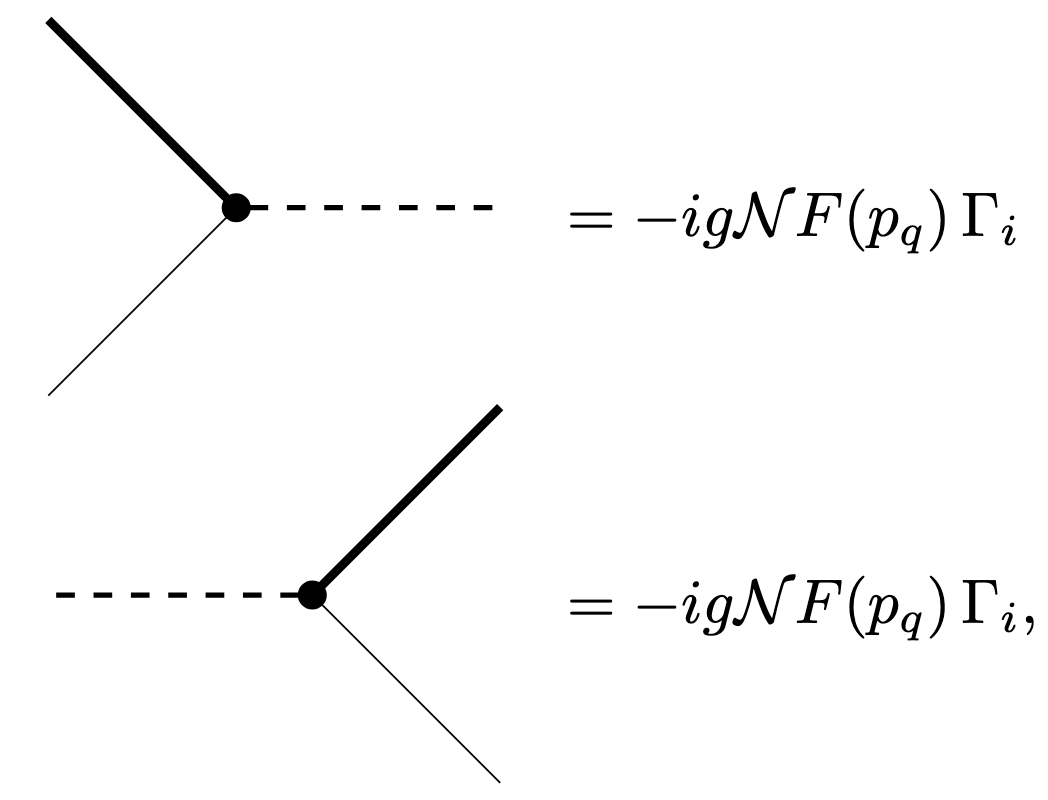}
    \caption{Vertices modified by instanton effects. $\Gamma_i$ are
      generically expressed as $\Gamma_{V} = -\gamma_\mu \epsilon_\mu$
      and $\Gamma_{P}=i\gamma_5$. Here, $\epsilon_\mu$ represents the
      polarization vector for the vector heavy meson.} 
    \label{fig:vertex}
\end{figure}

In the previous section, we presented the heavy-meson states and the
residual mass from the instanton vacuum. Here, we will use the
redefined vertex diagrams that incorporate the instanton effects, as
illustrated in Fig.~\ref{fig:vertex}. 
These modified vertices encapsulate the nonperturbative effects of the
instanton vacuum. In the limit of infinitely heavy quark mass, the
weak decay constants of heavy mesons are defined in terms of
axial-vector and vector currents: 
\begin{align}
    &\langle0|J_\mu^A|P\rangle = \langle0|q^\dagger \gamma_\mu
      \gamma_5h |P\rangle = -f_P \sqrt{m_P}v_\mu,\cr
    &\langle0|J_\mu^V|V\rangle = \langle0|q^\dagger \gamma_\mu h
      |V\rangle = if_V\sqrt{m_V}\epsilon_\mu,
    \label{eq:decay_cons}
\end{align}
where $|P\rangle$ and $|V\rangle$ denote the pseudoscalar and vector
heavy mesons, respectively. To determine the $1/m_Q$ correction to the 
decay constants, we consider the heavy-quark field $Q(x)$ defined in
Eq.~\eqref{HQF(x)}. We can express $Q(x)$ as 
\begin{align}
    Q(x)=e^{-im_Q v\cdot x}\left(1-\frac{1}{2m_Q}\Slash{D}_\perp +
  \mathcal{O}\left(m_Q^{-2}\right)\right)h(x), 
\end{align}
which leads to the following expressions for the vector (axial-vector)
current
\begin{align}
     J_\mu^{V(A)}(0)\simeq q^\dagger\gamma_\mu (\gamma_5) h -
  \frac{1}{2m_Q}q^\dagger \gamma_\mu(\gamma_5) \Slash{D}_\perp
  h+\mathcal{O}(m_Q^{-2}). 
\end{align}
Considering that the matrix elements also depend on $m_Q$, we must
account for the $1/m_Q$ corrections to the matrix elements given
in Eq.~\eqref{eq:decay_cons}~\cite{Neubert:1993mb}: 
\begin{align}
    \langle0| J_\mu^{V(A)} |V(P)(v)\rangle &=\langle0|q^\dagger\gamma_\mu
      (\gamma_5) h|V(P)(v)\rangle - \frac{1}{2m_Q} \langle0|q^\dagger
      \gamma_\mu (\gamma_5) \Slash{D}_\perp h |V(P)(v)\rangle \cr 
&- \frac{1}{2m_Q} \int d^4y
      \langle0| T \left\{q^\dagger\gamma_\mu(\gamma_5) h(0),h^\dagger
      iD_\perp^2h(y)\right\}|V(P)(v)\rangle+\cdots.
\end{align}
This expression is derived by modifying the heavy meson states as
follows: 
\begin{align}
    |\phi_v\rangle\rightarrow \left(1-\frac{1}{2m_Q} \int d^4x\, T\{
  \mathcal{L}_Q^{1/m_Q}(x)\}+\cdots \right)|\phi_v\rangle, 
\end{align}
where $\mathcal{L}_Q^{1/m_Q}$ is the next-to-leading order (NLO) term
of Eq.~\eqref{HQET_Lag}. The second term of $\mathcal{L}_Q^{1/m_Q}$
gives the spin-spin interaction term. 
After performing the instanton ensemble average, we obtain the
numerical results for the matrix elements of gluon operators as  
\begin{align}
    &\langle A_{I\mu}\rangle=0,\quad \langle G_{\mu\nu}\rangle=0, \quad
\langle A_I^2\rangle=\frac{24\pi^2\rho^2n N_c^2}{N_c^2-1}, \quad
      \langle A_{I4}^2\rangle=\frac{1}{4}\langle A_I^2\rangle,
\end{align}
where the instanton field $A_{I\mu}$~\cite{Diakonov:1989un} and the
strength tensor $G_{\mu\nu}$ are defined as 
\begin{align}
  A_{I\mu}(x)&=\eta_{\mu\nu a}^{(-)}
               \tau^a\frac{\rho^2x_\nu}{x^2\left(x^2+\rho^2\right)}, \;\;\;
  G_{\mu\nu}=\partial_\mu A_{I\nu}-\partial_\nu
              A_{I\mu}-i[A_{I\mu},A_{I\nu}] 
\end{align}
with the 't Hooft symbol $\eta^{(\pm)}_{\mu\nu a} = \epsilon_{a\mu\nu 
  4}\pm\delta_{a\mu}\delta_{4\nu} \mp\delta_{a\nu}\delta_{4\mu}$.

Since the heavy-light vector and axial-vector currents are not
conserved, we need to renormalize
them~\cite{Neubert:1993mb, Falk:1992fm, Neubert:1992fk, Falk:1990de,
  Manohar:2000dt}.  
\begin{align}
    J^{V(A)}_\mu = \sum_{i}C_i(\mu_0)\left(J_{\mu}^{V(A)(i)} -
  \frac{1}{2m_Q}\left(O_{\mu}^{V(A)(i)}+T_{\mu}^{V(A)(i)}
  \right)\right)+\mathcal{O}\left(\frac{1}{m_Q^2}\right),  
\label{eq:43}
\end{align}
where $C_i(\mu_0)$ denote the Wilson coefficients, expressed as 
\begin{align}
    C_1(\mu_0) = 1+\frac{\alpha_s(\mu_0)}{\pi} \left(\ln\frac{\mu}{\mu_0} -
  \frac{4}{3}\right),\qquad C_2(\mu_0) = -\frac{2\alpha_s(\mu_0)}{3\pi}. 
\end{align}
Here, $\mu_0$ is the normalization point from the current work. 
While, in the present
approach, the value of the normalization point $\mu_0$ is not uniquely
determined from first principles, we can set to be $\mu_0$
approximately by $\mu_0 \approx \bar{\rho}^{-1}$. We can examine the
plausibility of this value, following
Refs.~\cite{Diakonov:1995qy,Kim:1995bq}. We define $\mu_0 = a/\bar{R}$
with $\bar{R}^{-1} \approx 200\,\mathrm{MeV}$. Here, $a$ is a
dimensionless parameter varied from 3 to 7~\cite{Diakonov:1995qy} in
the variational estimate of bulk properties of the instanton medium.
$\mu$ is the renormalization point at the matching scale. 
$J_\mu^{V(A)(i)}$ in Eq.~\eqref{eq:43} represent the leading-order
currents defined as 
\begin{align}
    J_\mu^{V(A)(1)}=q^\dagger\gamma_\mu(\gamma_5) h, \quad
  J_\mu^{V(A)(2)} = q^\dagger v_\mu(\gamma_5) h. 
\end{align}
$O_\mu^{V(A)(i)}$ and $T_\mu^{V(A)(i)}$ stand for the currents arising from
the NLO in the $1/m_Q$ expansion  
\begin{align}
    &O_\mu^{V(A)(1)} = q^\dagger\gamma_\mu(\gamma_5)\Slash{D}_\perp
      h,\quad O_\mu^{V(A)(2)} = q^\dagger v^\mu(\gamma_5)
      \Slash{D}_\perp h,\cr 
& T_\mu^{V(A)(1)}=\int d^4y\left\{J_\mu^{V(A)(1)}
                               (0),h^\dagger(y)iD_\perp^2h(y)\right\},  
\quad T_\mu^{V(A)(2)}=\int d^4y\left\{J_\mu^{V(A)(2)}
      (0),h^\dagger(y) iD_\perp^2h(y)\right\}.  
\end{align}

\begin{align}
  \langle0&|J_\mu^{V(A)}|P(v)\rangle=-\mathcal{C}_{12}F_{V(P)}^{(0)} \boldsymbol{v}_\mu+
            \frac{\mathcal{C}_{12}}{2m_Q}(F_{V(P),1}^{(1/m_Q)}+F_{V(P),2}^{(1/m_Q)})
              \boldsymbol{v}_\mu =-\mathcal{C}_{12}F_P \boldsymbol{v}_\mu,\label{ax_cur_1/m} 
\end{align}
where $\mathcal{C}_{12}\equiv C_1(\mu_0)-C_2(\mu_0)$ and $\boldsymbol{v}_\mu$ defined as $\epsilon_\mu$ for the vector meson and $-iv_\mu$ for the pseudoscalar meson. 

We first compute the decay constants for the pseudoscalar and vector 
heavy mesons from the leading-oder currents.   
The results are obtained to be 
\begin{align}
    \sum_iC_i(\mu_0)\langle0|J_\mu^{A(i)}|P\rangle=& i\mathcal{C}_{12}g
    \mathcal{N}\int  \frac{d^4p_q}{(2\pi)^4} F(p_q) \Tr
     \left[\gamma_\mu\gamma_5S_Q(i\Lambda-v\cdot
   p_q)\Gamma_PS_q(p_q)\right]=-\mathcal{C}_{12}F_P^{(0)}v_\mu,
\label{decayP}\\ 
    \sum_iC_i(\mu_0)\langle0|J_\mu^{V(i)}|V\rangle= &
           i\mathcal{C}_{12}g\mathcal{N} \int 
  \frac{d^4p_q}{(2\pi)^4} F(p_q) \Tr
    \left[\gamma_\mu S_Q(i\Lambda-v\cdot p_q)
    \Gamma_VS_q(p_q)\right]=i\mathcal{C}_{12}F_V^{(0)}\epsilon_\mu,
\label{decayV}, 
\end{align}
where $S_Q$ and $S_q$ stand for the heavy-quark and light-quark
propagator in Fig.~\ref{fig:2}, respectively. The explicit calculation
yields the same results for $F_P^{(0)}$ and $F_V^{(0)}$ 
\begin{align}
    F_P^{(0)}=F_V^{(0)} = \frac{4\pi^2 g\mathcal{N}}{(2\pi)^4}
  \int_{0}^\infty dp_q F(p_q) \frac{p_q \left[p_q^2+2(\Lambda +
  M_q(p_q))(\Lambda -
  \sqrt{p_q^2+\Lambda^2})\right]}{p_q^2+M_q^2(p_q)}. 
  \label{eq:FP}  
\end{align}
As mentioned previously, the $N_f=1+1$ heavy-light quark interactions
does not remove the degeneracy due to the heavy-quark spin symmetry
even with $1/m_Q$ corrections. The improvement can only be achieved by
considering the $N_f=2+1$ interactions. We will leave it as a future
work. 

We evaluate the decay constants for the $D$ and $B$ mesons to be in
the range of 
\begin{align}
    f_D=&\frac{\mathcal{C}_{12}F_D}{\sqrt{m_D}}=(295.0-322.6)\,\mathrm{MeV},
        \;\;\;
    f_B=\frac{\mathcal{C}_{12}F_B}{\sqrt{m_B}}=(200.5-242.6)\,\mathrm{MeV}, 
\end{align}
where we have used the value of $\mu_0$ from 1.4 GeV to 0.6 GeV,
respectively. If we use $\mu_0=1$ GeV, we obtain the decay constants
as 
\begin{align}
    f_D=302.4\,\mathrm{MeV},\;\;\;
    f_B=209.8\,\mathrm{MeV}.  
\end{align}

We now compute the $1/m_Q$ corrections to the decay constants: 
\begin{align}
    F_{P,1}^{(1/m_Q)} =& g\mathcal{N}\int dp_q
 F(p_q)\left[\frac{p_q^3\left(\Lambda - \sqrt{p_q^2 +
  \Lambda^2} \right)}{2\pi^2[p_q^2+M_q^2(p_q)]}
 +\frac{p_q\Lambda\left\{p_q^2+2\Lambda
  \left(\Lambda-\sqrt{p_q^2+\Lambda^2}\right)
   \right\}}{4\pi^2[p_q^2+M_q^2(p_q)]}\right],
          \cr
    F_{P,2}^{(1/m_Q)} = &-g\mathcal{N}\int_0^\infty dp_qF(p_q)
                          \frac{p_q}{4\pi^2(p_q^2+M_q(p_q)^2)}\cr 
    & \times\Bigg[2\left\{p_q^2(3\Lambda - \sqrt{p_q^2+\Lambda^2}) +
      4\Lambda^2\left(\Lambda-\sqrt{p_q^2+\Lambda^2}\right)\right\}+3
      M_q(p_q)\left\{p_q^2+2\Lambda\left(\Lambda- \sqrt{p_q^2+
      \Lambda^2}\right)\right\}\Bigg]\cr  
    &+2g\mathcal{N}\frac{9\pi^2\rho^2N_cN}{(N_c^2-1)V} \int_0^\infty
      dp_q\frac{F(p_q) p_q \left(p_q^2-(2 \Lambda +M_q(p_q))
      \left(\sqrt{\Lambda ^2+p_q^2}-\Lambda \right)\right)}{4 \pi ^2 
 \left(p_q^2+M_q^2(p_q)\right) \sqrt{p_q^2+\Lambda ^2}}.
\end{align}

In Table~\ref{tab:1}, we list the total results for the heavy meson
decay constants, $f_D$ and $f_B$, obtained at the normalization point
given in the range of  $\mu_0=(0.6^{(+)}-1.4^{(-)})$ GeV. The results
are in good agreement with the PDG average
values~\cite{ParticleDataGroup:2024cfk}. We also compare the current
results with those from lattice QCD~\cite{FLAG:2021npn} and QCD sum
rules~\cite{Pullin:2021ebn}.     

%%%%%%%%%%%% TABLE %%%%%%%%%%%%
\setlength{\tabcolsep}{10pt} % Default value: 6pt
\renewcommand{\arraystretch}{1.6} % Default value: 1
\begin{table}[htp]
    \centering
    \caption{Total results for the heavy meson dacay constants and
      masses of the $D$ and $B$ mesons  in
      comparison with recent data from lattice QCD~\cite{FLAG:2021npn}
      and QCD sum rules~\cite{Pullin:2021ebn}. In the last column, PDG
      average values~\cite{ParticleDataGroup:2024cfk} are given. The
      normalization point in the present work is given in the range of
    $\mu_0=(0.6^{(+)}-1.4^{(-)})$ GeV.} 
    \label{tab:1} 
    \begin{tabular}{c|c|c|c|c}
          & This work & Lattice QCD~\cite{FLAG:2021npn} & QCD sum rule~\cite{Pullin:2021ebn} & PDG average~\cite{ParticleDataGroup:2024cfk}  \\ \hline\hline
$f_D\,[{\rm MeV}]$ & $204.6_{-4.5}^{+13.6}$ & $209.0(2.4)$ & $190(15)$
                & $203.8(4.7)(0.6)(1.4)$ \\ 
$f_B\,[{\rm MeV}]$ & $191.2_{-8.4}^{+30.0}$ & $192.0(4.3)$ &
  $192^{+20}_{-15}$ & $190.0(1.3)$ 
    \end{tabular}
\end{table}

%%%%%%%%%%%%%%%%%%%%%%%%%%%%%%%%%%%%%%%%%%%%%%%%
\section{Summary and Outlook}\label{summary}
%%%%%%%%%%%%%%%%%%%%%%%%%%%%%%%%%%%%%%%%%%%%%%%%
In this work, we aimed at developing an approach to heavy-light quark
systems based on the QCD instanton vacuum. We first
derived an effective heavy-light quark interaction for the $N_f = 1 +
1$ case, providing a connection between light and heavy quark
dynamics. By employing the compositeness and normalization
conditions, we determined the physical residual mass of heavy mesons
to be $\Lambda \approx 262.8$ MeV. This allowed us to calculate the
masses of $D$ and $B$ mesons, obtaining $m_D = 1.942$ GeV and $m_B =
5.322$ GeV, which approximate the average masses of pseudoscalar and
vector heavy mesons. We then computed the weak decay constants for the
$D$ and $B$ mesons to the leading order and next-to-leading order in
the $1/m_Q$ expansion. Considering the normalization point of the
current approach in the range of $\mu_0=(0.6-1.4)$ GeV, we obtained
$f_D = (204.6^{+13.6}_{-4.5})$ MeV and $f_B = (191.2^{+30.0}_{-8.4})$
MeV. The present results are in good agreement with recent lattice QCD
results and PDG average values. 
The current study establishes a foundation for further investigations
within the instanton vacuum framework. Future directions include
extending the analysis to the $N_f = 2 + 1$ case to examine spin-spin
interactions and their role in lifting the degeneracy between
pseudoscalar and vector heavy mesons. The heavy-light
quark interactions derived from the instanton vacuum may 
shed light on the nonperturbative structure of the heavy mesons and
illustrate the interplay between heavy quark symmetry and chiral
dynamics in more complex heavy-light quark systems such as tetraquark
hadrons.   

\section*{Acknowledgments}
The work was supported by the Basic
Science Research Program through the National Research Foundation of
Korea funded by the Korean government (Ministry of Education, Science
and Technology, MEST), Grant-No. 2021R1A2C2093368 and 2018R1A5A1025563
(NYGh and HChK).
\appendix
\section{Integration in color space}
\label{app:1}
In this appendix, we compile the formulae for the integration over
color orientation: 
\begin{align}
    &\int dU = 1,\qquad \int dU
              \,U^{\alpha_1}_{\beta_1}{U^\dagger}^{\alpha_2}_{\beta_2}
              =
              \frac{1}{N_c}\delta^{\alpha_1}_{\beta_2}\delta^{\alpha_2}_{\beta_1},\cr
   & \int dU U^{\alpha_1}_{\gamma_1}U^{\dagger
              \delta_1}_{\beta_1}U^{\alpha_2}_{\gamma_2} U^{\dagger
              \delta_2}_{\beta_2} = 
 \frac{1}{N_c^2}\delta^{\alpha_1}_{\beta_1}\delta^{\gamma_1}_{\delta_1}
 \delta^{\alpha_2}_{\beta_2}\delta^{\delta_2}_{\gamma_2} 
    +\frac{1}{4(N_c^2-1)}[\lambda^c]^{\alpha_1}_{\beta_1}
                                    [\lambda^d]^{\gamma_1}_{\delta_1}
                                    [\lambda^c]^{\alpha_2}_{\beta_2}
                                    [\lambda^d]^{\delta_2}_{\gamma_2}\cr    
    &= \frac{1}{N_c^2-1}\left[
      \delta^{\alpha_1}_{\beta_1}\delta^{\alpha_2}_{\beta_2}
      \left(\delta^{\delta_1}_{\gamma_1}\delta^{\delta_2}_{\gamma_2}
      -\frac{1}{N_c}\delta^{\delta_1}_{\gamma_2}\delta^{\delta_2}_{\gamma_1}
      \right) + 
      \delta^{\alpha_1}_{\beta_2}\delta^{\alpha_2}_{\beta_1}
      \left(\delta^{\delta_1}_{\gamma_2}\delta^{\delta_2}_{\gamma_1}
      -\frac{1}{N_c}\delta^{\delta_1}_{\gamma_1}
      \delta^{\delta_2}_{\gamma_2}\right)
    \right],
\end{align}
where Fierz identity for color matrices
\begin{align}
    [\lambda^c]^{\alpha_1}_{\beta_1}[\lambda^c]^{\alpha_2}_{\beta_2}
  &=     2\left(\delta^{\alpha_1}_{\beta_1}\delta^{\alpha_2}_{\beta_2}
    -\frac{1}{N_c}\delta^{\alpha_1}_{\beta_2}\delta^{\alpha_2}_{\beta_1}\right),
    \label{fierz-col-ll}
\end{align}
are used.
%%%%%%%%%%%%%%%%%%%%%%%
\bibliography{HL_mesonsIV}
\bibliographystyle{apsrev4-1}

\end{document}